# Empirical AI Ethics: Reconfiguring Ethics towards a Situated, Plural, and Transformative Approach

Paula Helm, *University of Amsterdam;* Selin Gerlek, *University of Amsterdam*

**Abstract**
Mainstream AI ethics, with its reliance on top-down, principle-driven frameworks, fails to account for the situated realities of diverse communities affected by AI (Artificial Intelligence). Critics have argued that AI ethics frequently serves corporate interests through practices of "ethics washing," operating more as a tool for public relations than as a means of preventing harm or advancing the common good. As a result, growing scepticism among critical scholars has cast the field as complicit in sustaining harmful systems rather than challenging or transforming them. In response, this paper adopts a Science and Technology Studies (STS) perspective to critically interrogate the field of AI ethics. It hence applies the same analytic tools STS has long directed at disciplines such as biology, medicine, and statistics to ethics. This perspective reveals a core tension between vertical (top-down, principle-based) and horizontal (risk-mitigating, implementation-oriented) approaches to ethics. By tracing how these models have shaped the discourse, we show how both fall short in addressing the complexities of AI as a socio-technical assemblage, embedded in practice and entangled with power. To move beyond these limitations, we propose a threefold reorientation of AI ethics. First, we call for a shift in foundations: from top-down abstraction to empirical grounding. Second, we advocate for pluralisation: moving beyond Western-centric frameworks toward a multiplicity of onto-epistemic perspectives. Finally, we outline strategies for reconfiguring AI ethics as a transformative force, moving from narrow paradigms of risk mitigation toward co-creating technologies of hope.

## Introduction

> *"Genuine hope is not blind optimism. It is a deliberate act of courage in the face of uncertainty."*
> Ernst Bloch, 1954

What distinguishes "good" from "evil" AI? How can we embed values into AI? Whose values should guide this process, and who should determine which values take precedence when tensions arise? These questions underscore the relevance of AI ethics as the field positioned to address them. However, mainstream AI ethics, in its current form, has been criticised for engaging in "ethics washing", a practice that reduces ethics to a tool for public relations and commercial decision-making. This approach ultimately serves the interests of big tech companies and their accomplices by mitigating risks, soothing public concerns, and reinforcing industry dominance while failing to promote the common good and prevent harm effectively. The perceived shortcomings of AI ethics have, in turn, sparked a wave of "ethics bashing" from critical scholars and civil society actors, many of whom turned deeply suspicious of AI ethics as a whole (Bietti 2019; Phan et al. 2022; Powell et al. 2022).

Taking these criticisms seriously, this paper takes a step back by beginning with asking: *How is AI ethics enacting values in relation to AI, where do its core problems originate from, and how can we address them?* In the first half of the paper, we hence make AI ethics itself the object of inquiry and investigate the origins, theory as well as practice of contemporary AI ethics. In doing so, we closely examine what has been described as "principlism" in ethics, a reliance on abstract, universal principles often accompanied by an unexamined, Western-centric essentialism (Clouser and Gert 1990). This trend is partly caused by industry-related research on operationalisable AI ethics. While not the sole cause, we follow the assumption that it is this approach, characterised by its lack of empirical grounding and its top-down reasoning—that has played a significant role in reducing AI ethics to a convenient tool for powerful tech corporations (Aradau and Blanke 2022; van Maanen 2022; Mager et al. 2025).

This is not to suggest that principlism alone is responsible for AI ethics' perceived ineffectiveness or its instrumentalisation by big tech. Structural factors also play a role — such as the expanding "economy of virtue" within the broader AI ecosystem, which turns ethical expertise into a commodity, or national austerity measures that force independent university researchers to rely on industry funding (Phan et al. 2022). At a deeper level, though, recent developments in AI, along with the myriad ethical challenges they bring, present a call to investigate the limitations of traditional rule-based approaches to ethics. While AI ethics has flourished as an academic field, it has accomplished astonishingly little in effectively addressing the real-world impacts of persistent ethical issues associated with AI. Despite extensive debates on fairness in AI, instances of proven algorithmic discrimination continue to emerge (Humber 2023). Privacy concerns have devolved into a state of privacy nihilism (Gertz 2024). Dystopian scenarios are increasingly becoming reality, with rising investments in autonomous weapon systems (Widder, Gururaja, and Suchman 2024). The troubling centralisation of power enabled by data-intensive, deep-learning AI has escalated with breakthroughs in transformer architectures (Luitse and Denkena 2021), while the energy demands of data centres have reached unprecedented levels (Lehuedé 2024b).

In this paper, we argue that it is precisely the field's historically *vertical* character–a reliance on rule-based methodologies–that has shaped AI ethics into a powerful but problematic framework. Its vertical character leaves AI ethics particularly vulnerable to co-optation for adversarial purposes, such as the legitimation of unequal power structures. Relying on narrow principles, AI ethics struggles to respond dynamically to the complex realities of various communities impacted by and appropriating AI technologies in diverse ways. The vertical approach not only risks overlooking the specificity of AI's harms in different domains and use cases. More importantly, it fails to account for the lived experiences and cultural nuances that shape how AI systems are implemented and perceived in different social contexts (Bonami et al. 2025).

Rather than taking the concepts and methods of AI ethics for granted, we subject them to the same critical scrutiny that Science and Technology Studies (STS) have applied for decades to disciplines such as biology, genetics, statistics, algorithms, medicine, and physics. STS has long examined how these fields co-produce the very phenomena they study: how population statistics construct categories of people (Hacking 1990), how biological taxonomies invent species classifications (Bowker and Star 1999), and how algorithmic predictions shape possible futures (Mackenzie 2015). We extend the same scrutiny to ethics. Like the scientific disciplines STS engages with, ethics is not a neutral arbiter of moral truth, but a historically and culturally situated practice. Its epistemic culture (Akrich 1992) co-produces the values, norms, and moral guidelines it often presents as universal, objective, and timeless.

This is particularly relevant in the field of AI ethics. From its inception, AI ethics has never spoken with a single voice. While vertical approaches have sought to impose fixed moral principles and rule-based frameworks onto AI, they have been accompanied by *horizontal critiques*. These horizontal approaches challenge universalist ethics in favour of context-sensitive, participatory, and relational methods. They emphasise the situated nature of ethical reasoning, recognising that values emerge from interaction, socio-political conditions as well as the inconsistent hopes, demands and desires people face in their daily lives. Yet, as we will show, they too come with limitations.

As a first step, this paper contributes to a de-essentialisation of AI ethics by situating it within the scholarly contexts from which it emerged. For this, the authors, trained in philosophy and anthropology, draw on their own experiences of working in applied ethics across various institutions and countries over the years, partly identifying with ethics as a field while never entirely fitting in. Reflecting critically on AI ethics as a field reminds us that 'things could be otherwise' (Haraway 1988). This perspective encourages the exploration of new pathways, recognising that embracing alternative practices and perspectives may help address some of the pressing challenges currently facing AI ethics. In the second part of this paper, we consequently move towards outlining three avenues for reconfiguring AI ethics. We start by setting the theoretical foundations: from top-down reasoning to empirical grounding. Then we move on towards a pluralisation: from western centrism to a multiplication of reference frameworks. Finally, we propose strategies for reconfiguring AI ethics as a transformative force: from risk mitigation to technologies of hope.

1. Situating AI Ethics

The historical narrative of AI ethics can be understood as a convergence of multiple scholarly traditions with significant influences from moral ethics, bioethics, data and information ethics, and the philosophy of technology. This convergence underscored the importance of conceptual clarity in defining ethical principles for emerging technologies. It is no surprise that ethical guidelines proposed for AI are largely modelled after broader frameworks in the ethics of technology–particularly those grounded in principle-based reasoning. While such frameworks have provided direction, they remain rooted in a *vertical approach*. As we will argue, this comes with severe problems that have already sparked *horizontal* critiques–emphasising the need for situated, practice-attuned forms of AI ethics that emerge from within design development and deployment contexts. Yet, these, too, leave us with serious limitations, which the second half of the paper then seeks to address.

**1.1. AI Ethics as a Historically Vertical Approach**

AI ethics guidelines largely follow the trajectory of general technology ethics, drawing on traditional fields such as bioethics, neuroethics, and medical ethics (Hagendorff 2020; Jobin, Ienca, and Vayena 2019). Luciano Floridi, among others, proposed adapting bioethics principles for AI by adding explainability to address the opacity of deep learning systems (Floridi et al. 2018). AI systems, particularly within generative AI, have gained prominence for their capabilities while remaining opaque 'black boxes (Zednik 2021; Rohlfing et al. 2021).' The lack of transparency raises pressing ethical and legal concerns around accountability, bias, and trust in AI decision-making. This challenge has not only shaped the agenda of AI ethics early on, but also helped define the broader contours of *data ethics* in the era of big data (Mittelstadt 2019; Taddeo and Floridi 2018).

This early focus has been foundational in shaping AI ethics through a rights-based approach inspired by moral philosophy (Fjeld et al., 2020). Key rights in these discussions include privacy, transparency, explainability, fairness, non-discrimination, and autonomy. While compelling in principle, these frameworks often fail in practice due to vague definitions, limited applicability, or weak enforcement. For instance, should privacy ever be traded off against security or efficiency? If privacy is understood as a fundamental right, or as essential to the realisation of another fundamental value like autonomy, such trade-offs would be ethically unacceptable (Rössler 2001). However, if privacy is perceived as a context-sensitive and normatively dependent value, it may be legitimately exchanged against competing values such as care, participation, or solidarity (Trepte, Scharkow, and Dienlin 2020).

This is not merely a problem of practical implementation, but one rooted in the disciplinary culture of philosophical ethics itself. Philosophical reasoning rarely yields consensus; values are neither self-evident nor universally defined. Depending on the ethical framework they adhere to, the socio-political positionality they bring to their work, and the specific cases they think with, philosophers often arrive at profoundly different–often incompatible–exegeses of the same concepts. Classic debates over negative versus positive freedom, and equity versus equality, illustrate how deeply contested even seemingly universal values can be. Consequently, application-oriented ethics must grapple with this plurality when seeking "the right way forward."

Ideally, this would require a deliberative process grounded in public reasoning and guided by the strength of arguments. But particularly in the context of AI, this ideal is often displaced by pragmatic consideration. What counts as the "best" ethical argument may depend more on technical feasibility, economic viability, or political expediency than philosophical rigour (Siffels and Sharon 2024; Helm 2024). Yet philosophical ethics continues to play a dominant legitimating role in AI ethics–and in public and corporate moral discourse more broadly–not for its practical use, but for its institutional authority and cultural prestige. As a discipline historically positioned to speak on moral truth, philosophy lends ethics discussions an aura of universality and neutrality. This status allowed it to frame the moral vocabulary of AI from the beginning in ways that appear objective and principled, masking the power dynamics and political contestation at stake. The result is an AI ethics that appears authoritative but is in fact abstracted from the messy conditions of realisation. Without questioning this, ethical debates risk reinforcing existing hierarchies – or worse – being reduced to a set of abstract values that can be invoked rhetorically while being selectively ignored in practice.

### 1.2. Corporate Co-Optation of Vertical Ethics
The gap between ethical commitments and real-world implementation is particularly evident in the corporate sphere, where major technology companies publicly endorse principles like fairness, sustainability, and explainability (Hagendorff 2020). Despite these declarations, they are often criticised for practices that contradict their professed values—such as unauthorised user data extraction (Zuboff 2019), biased AI systems (Buolamwini and Gebru 2018), exploiting data workers (Miceli and Posada 2022), or experimenting on vulnerable populations under the guise of humanitarian or democratic efforts (Madianou 2021). This pattern reflects what Sarah Ahmed calls "non-performativity": the act of announcing a commitment serves to shield organisations from scrutiny while allowing harmful practices to persist (Ahmed 2016). AI ethics thus risks becoming an exercise in performative reassurance, where ethical discourse functions more as rhetorical device than enforceable framework.

This persistent gap between ethical claims and practice underscores a fundamental flaw in vertical, top-down approaches to AI ethics (Fetic et al. 2020). Compounding this is a key structural issue: the lack of enforcement mechanisms for institutionalised AI regulations meant to address this gap. The EU's General Data Protection Regulation (GDPR, 2018), the Ethical Guidelines for Trustworthy AI by the European Commission's High-Level Expert Group (European Commission, n.d.), and the UNESCO AI Ethics recommendations (Hui 2023), represent institutionalisation efforts, but their impact has been inconsistent. The GDPR, for instance, aims to enhance individual control over personal data, yet struggles due to limited enforcement and reliance on overburdened public bodies (Goodman and Flaxman 2015). Even more fundamentally, its individualist framing can only partially address systemic threats to civil rights in the age of big data analytics. This moment calls for a reconfiguration of a long-established value: a shift from viewing privacy as liberal rights protection toward recognising it as foundational to democratic life across different global contexts (Helm and Seubert 2020).

Responding to legal shifts and public pressure, companies have created internal ethics teams or partnered with technical universities and state-funded centres. Conferences like the ACM Conference on Fairness, Accountability, and Transparency (FAccT) and the AAAI/ACM Conference on AI, Ethics, and Society (AIES) were established to foster interdisciplinary collaboration. These initiatives, originally grassroots efforts to challenge the power of big tech, have gradually been co-opted by industry (Phan et al. 2022). Today, major tech firms significantly influence the agendas of the critical community, using these venues not only to shape discourse but also as platforms for talent recruitment (Bruns 2019; Gebru and Torres 2024).

This corporate co-optation highlights a deeper issue with the vertical model of AI ethics, where ethical guidelines are often shaped by powerful corporate stakeholders rather than grounded in the lived experiences of affected communities. Vertical approaches have proven inadequate when faced with the complex power structures embedded in AI deployment. The tech industry's strategic influence and the absence of robust enforcement raise serious concerns about whether vertical AI ethics remains a meaningful guide for development or merely a tool for rhetorical commitment without accountability.

## 1.3. Horizontal Perspectives and their Limitations

In response to vertical, top-down approaches to ethics, horizontal, practice-oriented perspectives have emerged. These approaches trace their origins to broader critiques of the political dimensions inherent in technological artifacts. Technologies are not ahistorical, instead, they are situated within culturally and political contingent regimes of knowledge (Akrich 1992). This also applies to AI systems (Tacheva and Ramasubramanian 2023).

Transformer architectures, synthetic datasets, chatbots, robots, and wearables do not only mediate action, they enact specific epistemic assumptions, many of which are contested (Gerlek and Weydner-Volkmann 2025). Natural Language Processing development, for example, has long drawn on a Chomskyan notion of Universal Grammar, which assumes an underlying linguistic universality. This view, however, has been challenged by scholars who emphasise the profound and irreducible diversity of human languages (Evans and Levinson 2009). Ignoring such heterogeneity risks reinforcing linguistic homogenisation and reproducing colonial and imperialist legacies within multilingual modelling (Helm et al. 2024). While the rise of data-driven *Deep Learning* has moved AI beyond rule-based approaches, it has also exposed the limitations of *Universal Grammar*. Yet, although Deep Learning approaches have contributed to manifesting an astonishing extend of linguistic diversity, they remain skewed by the disproportionate dominance of English in training data. This imbalance surfaces in translation errors, especially where concepts lack English equivalents, as frequently visible in Google Translate (Bella et al. 2024). Treating AI as a mirror, this example highlights how a deeply rooted Anglocentrism, which is embedded in computer science as culture (Seaver 2017), is reflected in, and part of, scientific innovations.

If technologies reflect worldviews, then, in principle, they can be designed to advance certain values. Early frameworks such as *Value Sensitive Design* emphasise embedding ethics throughout the development process (Friedman 1997; Friedman and Hendry 2019; Knobel and Bowker 2011; van de Poel 2013). These proactive approaches aim to align innovation with societal values during early and midstream development (Fisher and Schuurbiers 2013). Tools like Explainable AI, fairness measures, dataset debiasing techniques, and the adaptation of red teaming for AI safety represent efforts to bridge principle and practice. As such, these approaches form part of a horizontal ethics that operates from within rather than imposed from above. Still, this raises key questions: who is included in this "within," who gets to decide, and whose voices remain excluded?

Despite being more implementation-oriented than vertical approaches, these methods often overlook power asymmetries and are shaped by design discourses that define ethical problems in technically solvable terms (Morozov 2014), such solutions may help, but often treat symptoms, not causes (Siffels and Sharon 2024). An example are predictive policing systems, which are trained on crime data to forecast behaviour instead of addressing root social causes and underlying social conditions through preventive measures like social work (Helm and Hagendorff 2021). This tendency reduces ethics to technical solutionism, addressing only those concerns that can be rendered manageable within existing technical frameworks.

The so-called *Moral Machine* experiment exemplifies another approach that, while appearing horizontal, still engages in reductive framing. Designed to explore the moral dilemmas of autonomous vehicles (Awad et al. 2018), it used a global online survey to gather input on who autonomous cars should prioritise in fatal scenarios. Participants chose between categories such as pedestrians versus passengers, young versus old, or humans versus animals. Though the study revealed cultural variation and underscored the value of diverse input, its methodology remained top-down: expert-led, based on constrained moral logics, and disconnected from the everyday experiences of those most affected by such technologies.

As AI ethics becomes increasingly institutionalised, shaped by governments, international organisations, and corporations, approaches centred on risk mitigation and design-level values prove insufficient. While initiatives like VSD and Moral Machine address some ethical issues, they tend to oversimplify how (un)ethical life unfolds in context. A more expansive ethical lens is needed, one that remains attuned to the social, political, and historical conditions in which technologies operate and evolve.

2. **Reconfiguring AI Ethics**

Despite these sobering critiques, we do not intend to join the chorus of AI ethics bashing. To do so would be to overlook the pervasive and performative power of moral judgments, which are inseparable from the politics of technoscience. Instead, our work takes up the question of "how to close the loop" in empirical ethics between situated observations and normative reasoning (Sharon and Koops 2021). We do this by outlining a three-pronged approach. The goal of this is to repurpose AI ethics and align it more closely with the concerns and aspirations of affected communities. The combination of strategies we are proposing aims to empower AI ethics as a transformative field, one that responds to communities needs and actively shapes AI rather than relegating it to a power-conserving, risk-mitigation role. The strategies are:
1. *Foundations:* Grounding AI ethics in empirical research to ensure care for the needs and concerns of diverse communities and domains impacted by AI.
2. *Pluralisation:* Multiplying frameworks of reference to invite diverse notions of ethics that allow for a move beyond Western-centric regulation.
3. *Transformation:* Embracing an ethics of possibility by engaging with alternative visions and material-discursive interventions leading to technologies of hope.

In the following sections, we outline each of them.

## 2.1. Foundation: From Top-Down to Situated Reasoning

Vertical approaches to AI ethics suffer not only from difficulties in practical application but also leave ethics vulnerable to co-optation by power. Their counterpart, horizontal approaches, in turn, often remain confined to risk mitigation and technical solutionism. The way forward, therefore, cannot simply be to replace vertical with horizontal ethics, but rather to recognise what critical perspectives on AI ethics share: an emphasis on empirical grounding and its consequences. This shift moves the analytical focus from principles to understanding what different actors and stakeholders consider desirable or undesirable practices—at individual, societal, and institutional levels. The challenge is then to connect these empirical insights to normative assessments of the good and the harmful (Hämäläinen 2017). In the pursuit of meeting this challenge, we draw from a range of disciplines and methodologies, including *Actor Network Theory* (Latour 1996), *Praxeography* (Mol 2002), *Postphenomenology* (Ihde 2009), *Everyday Ethics* (Pols 2015), and *Moral Anthropology,* (Fassin 2015), *Care Ethics* (Bellacasa 2017). These inductive approaches reject predetermined notions of what is inherently good or right (De Vries and Gordijn 2009). Instead, they invite a move from abstract critique toward analysing the development of relations between technologies, practices, and humans.

Emerging in the late 1990s, the Empirical Turn in philosophy marked a shift from dystopian critiques of philosophers like Heidegger and Anders, toward thinkers such as Don Ihde and Hans Achterhuis who emphasised technology's entanglement with everyday practices and institutions (Achterhuis 1992). This reflects a broader shift in the humanities and social sciences—from essentialist to relational accounts of technology. Related is the practice turn in the social sciences, which highlights routines and material engagements in shaping knowledge and norms (Cetina, Schatzki, and Savigny 2005). Thinkers like Pierre Bourdieu or Judith Butler reframed practices as key to understanding how bodies, technologies, meanings, and norms are co-constituted (Bourdieu 1977; Butler 1993). STS, within this lineage, pushed these turns further. Grounded in empirical research, stands like Actor-Network Theory and New Materialism offer three key insights: first, all phenomena are both epistemological and ontological, constituted through actor-networks (Barad 2007); second, agency emerges relationally through practice (Callon and Law 1997), and third, symbolic and technical mediations co-produce phenomena in situated ways. Reality, in this view, is not independent from observation or mediation (Callon 1984, Barad 2007).

Empirical philosopher Annemarie Mol, inspired by these insights, developed Praxeography – a methodology to examine how practices shape realities. Central to this is the concept of enactment: the idea that practices are socio-material and involve mutual agency between components (Mol 2002). These processes, through which different realities come into being, can be empirically observed and analysed. Mol extends Actor-Network Theory's symmetry principle to show how both human and non-human elements contribute to phenomena. Though her work focused on illness, it informs a de-essentialist view on AI, where coding, annotating, modelling, and prompting intersect with material infrastructures and assumptions about the human brain: how it learns and processes information.

Such a perspective carries profound implications for the practice of AI ethics. Like the horizontal approaches discussed earlier, it challenges the idea that data produced to train and fine-tune AI models simply mirrors reality; instead, it actively participates in shaping reality, as do the systems built upon it (Gitelman 2013). Moving further, Mol's approach is grounded in an onto-epistemic multiplicity that acknowledges the partial, situated, and contingent nature of both knowledge and embodied existence. This shifts ethical attention to tools and processes of observation themselves: how data is gathered, through which sensors or cookies, for what ends, which algorithms are deployed and based on what assumptions about truth and reality (see the example of Multilingual Models on p. 7/8).

Importantly, onto-epistemic multiplicity challenges the idea of universally applicable value concepts. Much like objects, worldviews, illnesses, or contested phenomena such as race (M'charek 2013), values are not pre-existing entities but are actively brought into being and stabilised through their enactment. This perspective contests essentialist notions of values, suggesting that although values may appear constitutive of functioning democracy and the good life, and their violations may cause real harm, they evolve through the very practices and interactions that performatively realise them as moral goods. Consequently, this understanding urges a reorientation of horizontal ethical approaches: from striving towards translating principles into practices, towards deriving principles from practices.

What, then, does it imply to derive principles from practices? It requires attending closely to everyday practices and recognising the diversity with which different people in different contexts interpret what constitutes ethical life, as well as how emerging value tensions might be addressed. Anthropologist Jeannette Pols, for example, has conducted extensive research on nursing practices, observing them as normatively oriented enactments of differing conceptions of what constitutes good caregiving and care-receiving, shaped by varying socio-technical conditions and positionings (Pols 2015; 2024).

These shifts in studying knowledge, ethics, and technology lay a foundation for reorienting AI ethics. Ethics is no longer about applying principles, but about engaging with socio-material practices. Within this intellectual landscape, Postphenomenology has emerged as a philosophical tradition that, while modifying the conceptual toolkit of STS, shares core assumptions. Just like STS, Postphenomenology takes embodiment seriously (Borgmann 2001; Ihde 2009). Focusing on situated encounters with specific technologies to investigate how they mediate human perception, action, and value-making, it aligns closely with the case-based approach found in STS and extends it with a phenomenological vocabulary for articulating human-technology relations. As Ihde argues, technologies can occupy the "of" in the phenomenological formula "consciousness of__" or "experience of__", thus becoming integral to the structure of experience itself (Ihde 2009).

From this perspective of a *mediated morality*, AI is not merely a neutral tool but a technomoral agent: it actively shapes moral experience and ethical judgment (Vallor 2016; Wellner 2024). By emphasising the role of practice and embodiment, postphenomenology helps understand how AI-driven technologies lead to the erosion of established values, such as the loss of public invisibility with the introduction of Google Glass (Kudina and Verbeek 2019). Conversely, technology can also acquire a moralising character. Postphenomenologist Peter-Paul Verbeek, for instance, has studied how the integration of predictive analytics with ultrasound can exert moral pressure on prospective parents (Verbeek 2008). Informed by such studies, Empirical AI Ethics recognises that socio-technological transformations continuously prompt the redefinition of ethical frameworks. Through diachronic analyses, it becomes clear how values may be *eroded*, *transformed*, or *rearticulated*, and new value commitments emerge through novel human-technology relations, as Friedrich et al. point out in their *eValuation* framework (Friedrich et al. 2022). This work helps de-essentialising ethics by revealing the historical and situated character of value concepts (Gerlek and Weydner-Volkmann 2025).

Another strategy for de-essentialisation lies in situating ethical life within distinct socio-cultural contexts. Consider, for instance, work in *Moral Anthropology*, such as that of Veena Das. Her ethnographies of violence survivors show how ethical life unfolds through everyday acts of care, resilience, and solidarity amidst trauma (Das 2012). This stands in stark contrast to sanitised thought experiments, like the trolley problem deployed in the Moral Machine, which fail to capture the embodied, messy, and situated nature of real-world ethical reasoning. These insights are crucial for AI ethics, where dominant frameworks often presuppose universal standards rooted in Western liberal ideals. As Das demonstrates in her study of post-violence survival among women in India (Das 2006), actions that might appear unethical from a Western perspective often emerge as pragmatic, necessary responses within local moral worlds. Such ethnographic sensitivities cultivate important insights for AI development. For example, credit scoring algorithms frequently misinterpret informal financial practices in the Global South as markers of risk, or corruption, overlooking how these practices can also embody acts of solidarity and care (Heeks and Swain 2018).

Such de-essentializing efforts may provoke questions regarding the status of normativity. Is normativity to be seen as completely relative to practice then? The answer is that while normativity doesn't vanish, it reorients. Judgment derives not from a standpoint of external reasoning but in relation to communities of practice. This analytic shift positions us well to describe what, following Boltanski & Thévenot might be termed *orders of worth*: the diverse justificatory logics that vary across communities of practice, domains, and individuals (Boltanski and Thevenot 2006). These varied orders of worth provide a grounded basis for formulating practice-informed repertoires of valuing. Often, what emerges from close-up observations and systematic analyses is a profound desire for transformation, manifested, for example, in the frustrations of nurses working under regimes and conditions that prevent them from enacting what they regard as 'good care' (Pols 2004). In such cases, existing orders of worth call for the critical revision of established value concepts.

Problematic consequences of the power-preserving character of rule-based ethics have already been documented in the context of AI. One infamous example regards the use of decision-support systems trained on extensive historical crime data to estimate the likelihood of reoffending, thereby influencing parole decisions. In such high-stakes applications, it is crucial that systems do not reinforce discriminatory patterns. However, studies have shown that popular fairness metrics based on a classical Rawlsian conception of 'justice as fairness' (Rawls 1970) often fail to account for intersectional inequalities (Barocas, Hardt, and Narayanan 2023). Instead, they may inadvertently sustain existing hierarchies. In practice, they have, for instance, proven insufficient in preventing disproportionately harsh treatment of offenders with darker skin shades (Lum and Isaac 2016). Notably, qualitative investigations have revealed irritations not only among those who felt unfairly targeted but also among those ostensibly advantaged by the system (Angwin 2015).

An empirically grounded AI ethics, one that begins with the systematic observation of the practices, perspectives, and lived consequences experienced by affected actors—may call for the formulation of new value concepts or the critical refinement of existing ones. For instance, systems previously evaluated through utilitarian or procedural notions of justice as fairness might be reassessed in light of historically situated conceptions such as epistemic justice (Helm et al. 2024; Kay, Kasirzadeh, and Mohamed 2024). This reorientation can also involve extending domain-specific values into new territories, such as the application of care, traditionally confined to reproductive labour and healthcare, to fields like law enforcement (Asaro 2019) or environmental protection (Bellacasa 2017).

Care ethics, with its relational understanding of moral life, aligns closely with the project of Empirical AI Ethics. As Joan Tronto famously stated:

> "Care is a species of activity that includes everything that we do to maintain, continue, and repair our 'world' so that we can live in it as well as possible. That world includes our bodies, our selves, and our environment, all of which we seek to interweave in a complex, life-sustaining web (Tronto 1998, p. 40)."

Rather than grounding itself in abstract principles or rights-based frameworks, care ethics centres on experience and the moral urgency of attending to the needs of others—especially those rendered vulnerable or marginalised. By placing repair at its core, care ethics also foregrounds the historical embeddedness of ethical life, recognising that before questions of procedural justice can be meaningfully posed, there is often a demand to redress the accumulated weight of past and ongoing harm. In resonance with STS and Postphenomenology, care ethics also emphasises the co-emergence of human subjects, technologies, and values, entangled in concrete, and frequently unequal, relational worlds.

What, then, would it mean to evaluate an AI system in criminal justice not through the logic of distributive fairness, commonly quantified via the proportional distribution of false positives, but through the *logic of care*? Such a shift would reorient our focus: from abstract metrics to the situated enactment of historically sedimented vulnerability. It would prioritise the repair of fractured social worlds, worlds in which trust in the criminal justice system has, for many, been repeatedly and systematically eroded by lived experiences of mistreatment, often based on race, gender, or other markers of social difference. This erosion of trust is not incidental; it feeds into cycles of recidivism and violence, perpetuating the very forms of marginalisation that AI systems purport to neutrally assess.

From a care ethics perspective, addressing algorithmic bias is not a matter of tuning models or refining datasets. It becomes a demand to confront and transform the social and historical conditions that produce inequality in the first place. It challenges us to resist the allure of technocratic solutions that deflect responsibility, and instead to engage in the difficult, relational work of repair.

## 2.2. Pluralisation: From West Dominance to Decentralisation

Grounding ethical reasoning in practice-based methodologies and research opens the possibility for ethics to function as a field responsive to everyday realities, rather than one operating merely top-down. Yet for Empirical AI Ethics to become transformative in a broader sense, responsiveness at the micro level must be connected to larger scales: the ways ethical concerns are institutionalised, regulated, and contested across different domains and publics. Addressing this complexity demands that ethical inquiry be pluralised—by critically reflecting on the limitations of its Western-centrism and by confronting the tensions between attentiveness to situated practices and the demands of regulating AI.

As outlined above, where empirical research articulates a clear demand for transformation, it may be most fruitful to seek inspiration in contexts that are most different from those in need of change. A plurality of conceptions of the good life flourishes beyond the confines of dominant Western principles of fairness, accountability, and transparency. Amerindian notions of Sumak Kawsay (*buen vivir*), for instance, foreground the entanglement of human and more-than-human actors by emphasising conditions of living well in reciprocal relation with others and with the land (Acosta 2013; Bonami et al. 2025). Similarly, *Ubuntu* ethics enable a departure from utilitarian logics, privileging interconnectedness, solidarity and collective prosperity as fundamental preconditions for individual well-being (Makulilo 2012; Mhlambi and Tiribelli 2023). In addition, *cosmopolitical approaches* invite a further expansion of Western conceptions of universal values by acknowledging that different ontological commitments—regarding the nature of reality and humanity's place within it—fundamentally shape the political and regulative processes through which societies define and pursue what is most important to them (Ingram 2013; Stengers 2011).

   Despite their varied genealogies, and akin to methodologies such as Actor-Network Theory, New Materialism and Postphenomenology, these orientations share a rejection of anthropocentric and dualistic models of agency. Instead, they are grounded in epistemologies that assume an infinite regress of relationality, where no actor, human or otherwise, exists in isolation. When applied to the domain of AI, such perspectives invite a more expansive ethical gaze that foregrounds the impact of technological systems on more-than-human well-being. This shift reframes environmental concerns not merely as technical issues of sustainability, but as deeply entangled with broader questions of human and ecological flourishing. Within this worldview, any AI innovation that produces environmental harm cannot plausibly be considered ethically desirable, rendering cost-benefit analyses as logically and morally untenable, even where short-term individual gains may appear to accrue.

   At the same time, these perspectives on more-than-human interconnectedness reveal a tension between situated ethical reasoning and the vast scale and reach of contemporary AI technologies. While it is essential to ground ethics in situated practices, we must also grapple with the undeniable need for coordinated responses to the global implications of AI. The substantial energy demands of large, multimodal models underscore the urgency of this challenge (Lehuedé 2024a). Their widespread deployment across multiple sectors necessitates, on various levels and for diverse reasons, the establishment of regulatory processes aimed explicitly at safeguarding the flourishing of plural forms of life (Hecht 2018). While perspectives focused on the micro-level of situated practices may on first sight seem ill-equipped to address large-scale systems and opinion formation, it would be equally problematic to retreat from these arenas. We must not cede this space to Western-centric rule-based models that reduce socio-technical complexity to quantifiable threats (Amoore 2023). As posthumanist Katherine Hayles aptly asks: how can we move beyond narrow, anthropocentric understandings of agency while preserving some of humanity's most significant ethical achievements, such as the *Universal Declaration of Human Rights* (Hayles 2017)?

What is needed are approaches that are capable of addressing different scales. Central to this endeavour is the recognition that ethical life can find its expressions materialised through *assemblages:* situated and emerging formations that link locality to wider networks of responsibility and meaning. One promising direction for cultivating such assemblages comes from the notion of *pluriversality*, inspired by the Mexican Zapatista movement and theorised by Arturo Escobar (2020). Pluriversality invites us to envision "a world where many worlds fit." Rooted in decolonial and post-development thought, it creates space for diverse ways of knowing, being, and doing. Inherent in the notion of pluriversality is a sharp critique of the universalising tendencies of contemporary ethics, even within well-meaning horizontal practices of co-creation, which can inadvertently reproduce colonial patterns by exporting Euro-American values under the banner of participation (Helm et al. 2023). In contrast, pluriversality supports autonomous designs of sociotechnical assemblages: forms of making and thinking that emerge from within communities, grounded in local concerns, cosmologies, and lifeways.

It is this shift of gaze that animates the approach proposed here: moving away from the assumption of an allegedly universal canon of values toward an inquiry into how diverse onto-epistemic positions can shape what is considered good or harmful AI, inspiring new modes of reasoning along this continuum. With its planetary perspective yet concern for the protection of local autonomy, pluriversality offers a compelling normative orientation for dealing with the ambiguity sketched out above (Escobar et a. 2025). It foregrounds the coexistence of multiple ways of being, knowing, and valuing across societies, while acknowledging the material-discursive interdependencies of ecosystems and infrastructures.

A second avenue for configuring the formation of assemblages lies in and is articulated in light of the Anthropocene through the lens of *interscalar vehicles* (Hecht 2018). In this framing, scale is not a pre-given metric of size or reach, but an emergent, performative effect of technopolitical, social, and cultural practices (Edwards 2002). Interscalar vehicles shape and are shaped by the categories we choose, the orientation we adopt, and the comparisons we authorise. In this sense, they are not metaphors, but instruments for ethical coordination and intervention across domains (Dijstelbloem 2021). They facilitate coordination without collapsing the local into the global or vice versa. Together with pluriversality, they foreground emerging assemblages of concern that maintain ways of knowing, acting, and valuing across shifting scales and sites.

> In science-fiction dreams of interstellar travel, characters travel distances unbridgeable by conventions of Newtonian mechanics. They arrive at impossible destinations, worlds that teach them new ways of seeing and being. Let's attempt similarly impossible journeys. What happens when we treat empirical objects as interscalar vehicles, as means of connecting stories and scales usually kept apart? (Hecht, 2018, p. 115)

Such vehicles are at work in the formation of transnational assemblages that seek justice across fragmented legal regimes, such as when former mineworkers in Gabon bypassed state mechanisms and sought recourse through NGOs in France and Niger to demand justice for uranium-induced illnesses and to make visible the slow violence of *toxic infrastructures*. Interscalar vehicles can operate via international standards, which serve both as the grounds and instruments of intervention and resistance. When successful, such assemblages perform interscalar work, comparing disparate procedures against shared benchmarks.

Philosopher Huub Dijstelbloem (2021) deepens this insight through the concept of infrastructural investigations, which expose and reassemble controversial structures of power. He examines the theories and practices by which journalists, activists, artists, and investigative organisations contest diffuse forms of power. Using techniques such as mapping, counter-databases, interactive visualisations, and testimonials, these actors reconfigure relations of visibility and accountability. They counter forms of organised irresponsibility, particularly evident in domains such as border politics, journalism, and supply chains, where harm is systemic yet dispersed, and difficult to attribute, a situation exacerbated by the opacity of AI black box technologies. Whether instantiated through crude AI quality benchmarks (Raji et al. 2021), algorithmic news recommenders (Harambam, Helberger, and van Hoboken 2018), or habit-forming interfaces (Ash et al. 2018), exposing socio-technical vehicles as forms of power operating within broader ecosystems of influence challenges linear notions of causality and, by extension, top-down governance (Gerlek and Dijstelbloem 2025). Making these dynamics legible as assemblages of concern can foster a shift from vertical ethics to an ethics of co-responsibility, reframing agency not as possession, but as circulation.

Notions of pluriversality, interscalar vehicles or infrastructural investigations emphasise situated practices without collapsing into relativism. They do not negate the shared responsibilities arising from interdependence. In a world of entangled infrastructures, precarious ecologies, and governing multi-purpose AI models, regulation demands more than policy; it requires onto-political awareness. The idea of publics as *coming into being*, theorised by John Dewey and Bruno Latour, is instructive in this context. Here, publics are not pre-existing entities but emerge through controversy, breakdown, and shared matters of concern (Latour and Weibel 2005), ultimately building assemblages of concern. Technologies such as AI, however, do not merely mediate communication or disseminate information, they actively participate in the assembly of publics by shaping what is rendered visible, debatable, or ignored (Gerlek and Dijstelbloem, 2025). If we assume that publics play a role in shaping regulation, and that AI is simultaneously co-constituting the very publics from which regulatory consensus is meant to emerge, we risk a recursive paradox: the goat becomes its own gardener. Assemblages draw attention to these new configurations of cross-scale interventions.

Addressing large-scale infrastructural injustices requires nurturing public assemblages of concern as meso-level formations. Emerging within and across domains such as environmental law, digital governance, and humanitarian logistics, these assemblages create spaces where partial, pluriversal, and situated practices can articulate across broader scales without flattening diversity or collapsing into universality. Rather than presupposing fixed domains, attention to emerging assemblages foregrounds the lived ways in which socio-technical practices, infrastructures, and publics coalesce. By enabling cross-scalar coordination, such assemblages play a crucial role in enacting a co-responsibility that is attuned to the complexities of a pluriversal world.

**2.3. Transformation: From Risk-Mitigation to Technologies of Hope**
We have discussed how we might attend to the situated, embodied character of ethical life, generating varied orders of worth across distinct domains, cultures, and communities of practice. We have also acknowledged and considered ways to deal with an ambiguity that stretches between the need for more overarching regulation and the aspiration to nurture pluriversal flourishing. As a third path, we will now highlight strategies for mobilising AI as part of enacting Empirical AI Ethics.

Rather than being driven by a relentless pace of innovation, where damage prevention dictates the agenda, AI ethics should contribute to manifesting alternative possibilities, possibilities that may appear improbable yet possess transformative potential (Leonelli 2024). Just as the historical development of STS evolved from critique into an engaged, interventionist programme (Latour 2004; Sismondo 2017; Latour and Weibel 2005), Empirical AI Ethics should not exhaust itself in analysis alone but venture into doing, into critical tool- and world-making (Ratto 2011). STS has demonstrated the capacity to articulate activating messages and to propose alternative visions and practices for confronting societal challenges (Haraway 2017). Our vision for Empirical AI Ethics follows this trajectory. While we acknowledge the legitimacy of despair in the face of devastation (Foer 2019), it represents a conscious ethical choice not to begin from that place, but from a firm belief in the generative power of concrete hope.

Amidst all contextualisation, certain foundational commitments remain that most would recognise: if harm or destruction is observed, there is an ethical imperative to respond. Yet before action can emerge, there must first be a sensation, a representation, an arousal that catalyses the will to act, and from this, hope is born (Willow 2023). Not the dulled optimism that numbs into inaction (Berlant 2011) but a hope that energises: a tangible, grounded form of hope. The kind that Ernst Bloch (1954), still haunted by the cruelties of the Second World War, termed *concrete hope* (Bloch 1985). To conceptualise hope as concrete, as something realised through practice, allows us to distinguish between two fundamentally different orientations. One is a hope suspended in abstraction: a projection onto distant, often ill-defined futures that remain too remote to motivate action. The other is a concrete, situated form of hope, one that emerges from feasible, local alternatives rooted in everyday needs. This latter form is neither naïve nor utopian; rather, it acknowledges conditions of constraint and even powerlessness (Han-Pile 2017).

This orientation towards hope, from trying to predict what is likely to happen and mitigating associated risks, towards attending to the unlikely but possible, resonates with Arjun Appadurai's distinction between an 'ethics of probability' and an 'ethics of possibility' (Appadurai 2013). The former limits us to the logic of what is most likely, a logic that binds us to focus on the visions of those who are in power (Appadurai 2013, p. 188, 295). The latter, in contrast, opens spaces for imagination and expands our collective capacity to aspire. Favouring the latter, Appadurai posits that an ethics of possibility 'can offer a more inclusive platform for improving the planetary quality of life and can accommodate a plurality of visions of the good life' (2013 p. 299–300).

Inspired by this vision, Empirical AI Ethics may begin to look beyond the well-worn terrain of probability, opening itself to pluriversal figurations of technology. An example comes from a project in which members of the Guarani-Kaiowá and Ikpeng-Xingu, Indigenous peoples of the Upper Amazon, were invited to imagine technologies they desired. Many participants responded with paintings of mythical beings: animals with human traits. At first, the facilitators assumed the prompt had been misunderstood. But the images were intentional. Among them was the *Nyra Nhe'Éngatu*, a radiant hybrid of bird and human, drawn from Ikpeng cosmology. This creature bridges earth and sky; a guardian of the unseen, responsible for the care of communication itself. It is said to visit those in need of sending urgent messages, especially in moments of injury or danger. One twelve-year-old boy painted this being to represent what he believed a smartphone should do for his people (Bonami et al. 2025).

Unlike Western AI benchmarks, which are typically governed by anthropocentric ideals or operational efficiency, the *Nyra Nhe'Éngatu* functions as an onto-epistemic opening (Cadena 2015): an orienting force that resonates closely with ideas of care ethics while diverging fundamentally from anthropocentric frameworks of explainability, accountability or fairness. Through collaboration with local communities of practice to understand their situated visions and needs, an *Empirical AI Ethics* approach has the potential to challenge dominant Silicon Valley narratives that valorise human-likeness as primary indicators of successful AI innovation. This dominant model presumes the inevitability of centralised, resource-intensive, and infrastructure-heavy AI systems (Brown et al. 2020). In contrast, the example discussed above offers an alternative vision for AI, one imagined not to replicate human capacities for convenience, but to assist in tasks that humans are inherently unable to perform, all while remaining within deeply humane registers of worth: supporting and caring for those in need.

By spotlighting examples that emphasise possibilities rather than probabilities, we seek to make a case for moving away from mere reactionary responses to technological advances, risks, and harms (Appadurai, 2013). An Empirical AI Ethics that seeks to be transformative should focus on elevating the hopes, desires, and aspirations of otherwise marginalised and unheard people. This helps to move away from a mere critical perspective on AI as a "naughty child" needing discipline; instead, AI can and should be part of the solution within this approach. And indeed, across the globe, numerous initiatives illustrate how data science can be harnessed to empower marginalized communities, for restorative justice, to challenge oppressive systems, and foster domain specific values. In the following we will give an example of each of these attempts to empirically enact normative visions through the help, adjustment, invention and further development of technical tools.

The *Life Language Initiative* serves as an example of a technology of hope that strives to elevate marginalised voices. This project functions as a database, catalogue, and collaborative platform, surpassing existing language databases not in volume, but in the diversity of languages represented. Its name – Live Language – reflects the fact that languages are alive; they are born, evolve, and die, along with the people who speak them (Giunchiglia et al. 2023). While most language databases focus on the 40 most resourced languages, thereby pivoting around English, Live Language seeks to document the linguistic diversity of the 6,960 remaining languages and dialects spoken today. This endeavour is vital for ensuring diversity on the internet. The vision is to collaborate with marginalized speakers around the world to create a pluriversal alternative to centralised models, fostering the development of tailored alternatives, a world where many languages fit (Koch et al. 2024).

The *Counting Femicide Movement* is another powerful example of how data science can contribute to reparation through the restoration of justice (D'Ignazio 2024). Counting Femicide is a collaborative initiative aimed at documenting and analysing gender-based killings of women and marginalised individuals worldwide. It addresses the systemic failure of governments to accurately track femicide by collecting data through grassroots efforts, media reports, and community input. The project seeks to make femicide visible by providing a comprehensive, publicly accessible database that highlights patterns of violence and state inaction. By leveraging data science methods and storytelling, initiatives like this aim to raise awareness, demand accountability, and advocate for policy changes to combat gender-based violence globally.

Other initiatives focus on efforts to enact domain-specific orders of worth, thereby countering the top-down tendencies associated with the algorithmisation of the public sphere (Aradau and Blanke 2022). This approach resonates with the earlier discussion regarding the necessity of assessing the enactment of values differently across distinct domains, where diverse communities of practice may embrace different orders of worth, each accompanied by distinct repertoires of valuing. To foster one such repertoire, *cultural citizenship*, which has been established as a key order within the cultural journalism sector, Ferraro et al. propose and, in fact, prototypically enact four concrete steps for the redesign of recommender algorithms for music curation: 1. Moving beyond a purely commercial orientation. 2. Recognising the cumulative influence of recommendations. 3. Integrating a metric of commonality that measures how well recommendations familiarise a given user population with specific content categories. 4. Avoiding a solely top-down normative approach by complementing it with experiments that assess the effectiveness of commonality in achieving the target value (Ferraro et al. 2024). This reconfiguration of algorithmic recommendation in the music domain is a prime example of enacting an ethics of possibility to create technologies of hope.

**Conclusion**
Mainstream AI ethics remains largely anchored in vertical, rule-based frameworks which frequently falter when confronted with the complexities of real-world application. Rooted in Western epistemologies, these approaches fail to account for the textured, lived realities of the diverse communities affected by AI technologies. In doing so, they risk reinforcing the very hierarchies and exclusions they are supposed to challenge. More troubling still, the focus on rule-based ethics and technical "solutions" has left AI ethics susceptible to co-optation by powerful industry actors, reducing it to a tool of risk management rather than allowing it to flourish as a transformative force.

In response to these limitations, we propose a shift toward empirically grounded, pluralistic approaches to AI ethics. Drawing on insights from STS, Empirical Philosophy, Postphenomenology, and Anthropology, we suggest that ethical values are not static, universal truths to be applied, but emergent, enacted through practices, shaped by specific cultural, historical, and material conditions. Findings from research rooted in these traditions disrupt essentialist conceptions of ethics and call for approaches attentive to the onto-epistemic multiplicity of ethical life as it unfolds across domains and communities of practice.

Such an approach insists on the importance of engaging with those most affected by technological systems, of studying their enactments of values, their performative concepts, their "orders of worth." These insights should not be seen as peripheral but as foundational for shaping regulatory frameworks that aim to be just and generative. To escape the narrow confines of a paradigm fixated on risk mitigation, we call for a reorientation toward possibility. In keeping with a practice-based ethics, we advocate for an ethics of hope grounded in the concrete, a commitment to making the possible liveable through situated, normatively oriented actions.

An empirically grounded ethics of hope draws inspiration from many sources: not only scholarly canons but rituals, daily routines, ecosystems, tools, and companion species. Ethical life, after all, does not begin in abstract principles but in lived experience. To engage with this breadth, empirical ethicists mobilise a rich repertoire of methods—from participant observation and creative storytelling workshops to the design and curation of databases, archives and algorithms. Through this work, they gather what Sarojini Nadar has called "data with souls", forms of knowledge that carry with them the resonance of human vision, pain, joy, and hope (Nadar 2014). These are not merely data points, but fragments of life: gestures, stories, spontaneous reactions, semiotically charged artifacts that refuse to be flattened into metrics.

Such data also remind us that many of the world's languages and cultures arise not from written scripts but from oral traditions. Much of what has endured across generations has done so without technological scaffolding, preserved not in code and energy intensive data centers, but in bodies, voices, memorised stories. This reminder offers a vital insight into our contemporary entanglements with AI. The belief that intelligent machines and digital archives are necessary to safeguard cultural knowledge is, at its core, an ideological assumption. It is part of a techno-solutionist worldview that positions AI as the inevitable vessel of human progress.

But the endurance of oral traditions teaches us otherwise: it is not technology that ensures the survival of meaning, it is care. What communities cherish, they nurture, through relationships, rituals, and remembrance. No algorithm can substitute this fidelity. As María Puig de la Bellacasa reminds us, care is not a soft supplement to knowledge but it's very condition (2017). This reorientation challenges the problem-solution logic that underpins much of AI ethics, even in its more horizontal forms. It invites us to privilege meaning over metrics, relationality over optimisation. In this light, Empirical AI Ethics is not merely a tool for resolving ethical dilemmas; it becomes a way of cultivating connection, between people, traditions, technologies, and the more-than-human world we inhabit.